\documentclass{ws-procs9x6}

\newcommand{\be}{\begin{equation}}
\newcommand{\ee}{\end{equation}}
\newcommand{\beq}{\begin{eqnarray}}
\newcommand{\eeq}{\end{eqnarray}}

\usepackage{graphicx}%
\usepackage{epsfig}

\begin{document}

\vspace*{-2.8cm}

\title{$N$ to $\Delta$ electromagnetic and axial form factors in full QCD}

\author{C. Alexandrou}

\address{Department of Physics, University of Cyprus,
P.O. Box 20537, 1678, Nicosia, Cyprus\\
E-mail: alexand@ucy.ac.cy, 
www.ucy.ac.cy}

\vspace*{-0.3cm}



\begin{abstract}
Lattice results on the N to $\Delta$ electromagnetic, axial-vector
and pseudoscalar form factors are evaluated
using dynamical staggered sea quarks
and domain wall valence quarks for pion masses in the range of 580-350 MeV,
as well as, dynamical and quenched Wilson fermions for similar pion masses.
\end{abstract}

\vspace*{-0.2cm}

\keywords{Lattice QCD, Form factors, Nucleon Resonances}

\bodymatter

\vspace*{-0.1cm}

\section{Introduction}\label{sec:Introduction}
State-of-the-art lattice QCD calculations can yield model independent
results on N to $\Delta$ transition form factors,
thereby  providing direct comparison with experiment.
One such example is the N to $\Delta$
quadrupole form factors that have been accurately measured 
in a series of recent 
experiments at low~\cite{Bates:1999,Mainz}
 and high momentum transfers~\cite{CLAS:2001}.
They encode information on the deformation of the nucleon and $\Delta$.
We present results on  these N to $\Delta$ electromagnetic
form factors, as well as on the dominant
axial-vector 
N to $\Delta$ transition form factors $C_5^A(q^2)$ and $C_6^A(q^2)$.
Experiments using electroproduction of the $\Delta$ resonance
 are in the progress~\cite{G0:2002} 
to measure
the parity
violating asymmetry in N to $\Delta$, which, to leading
order, is connected to 
$C_5^A(q^2)$.
Evaluation of the pseudoscalar 
$\pi N\Delta$ form factor, 
 $G_{\pi N\Delta}(q^2)$,  
follows once the N to $\Delta$
sequential propagators are computed.
In addition, we evaluate  the nucleon axial-vector
form factors and the $\pi N N$ form factor,
 $G_{\pi N N}(q^2)$. Having both the nucleon and the N to $\Delta$ 
form factors allows us
to discuss ratios of form factors  
that are expected to show
weaker quark mass dependence
and be less sensitive
to other lattice artifacts.
Furthermore, knowledge of the axial-vector form factors 
and the $\pi NN$ and
$\pi N \Delta$ form
factors allows us to check the Goldberger-Treiman relations.

 The light quark regime is  studied in two ways:
 Besides using  configurations with
two degenerate flavors of dynamical Wilson fermions
  we use a hybrid combination of domain wall valence quarks, which have chiral
symmetry on the lattice, and
MILC configurations generated with three flavors of staggered sea
quarks  using the Asqtad improved action~\cite{MILC}.

\vspace*{-0.3cm}

\section{Lattice Techniques}\label{sec:Lattice}

Observables in lattice QCD 
are given by the vacuum expectation value of gauge invariant
operators in Euclidean time:
\be 
<\Omega|\hat{O}|\Omega> =\frac{1}{Z}\int d[U]d[\bar {\psi}]d[\psi]\>\> 
O[U,\bar{\psi},\psi] 
e^{-S_g[U]-S_F[U,\bar{\psi},\psi]} 
\ee
Integrating over the fermionic degrees of freedom
we obtain
\be 
<\Omega|\hat{O}|\Omega> =\frac{1}{Z}\int d[U] \>\>\det(D[U])
O[U, D^{-1}[U]] 
e^{-S_g[U]}
\label{EV}
\ee
where $ D^{-1}_{jn}[U]$ substitutes each appearance of
 $-\bar{\psi}_n \psi_j$ in the operator and describes  valence quarks
whereas  $\det(D[U])$ corresponds to sea quarks.
The path integral over the gauge fields is done
 numerically by stochastically 
generating a representative ensemble
of gauge configurations according to the probability \vspace*{-0.5cm}
\be
P[U]=\frac{1}{Z}\exp\left \{-S_g[U]+\ln\left(\det(D[U])\right)\right \}.
\ee 
In this work, besides Wilson fermions for the sea and valence
quarks, we use staggered sea quarks ($\det(D_{\rm staggered}[U])$) and
domain wall valence fermions ($D^{-1}_{\rm DW}[U]$).
The expectation values are obtained by summing over the $U$-ensemble:
$<\Omega|\hat{O}|\Omega>=\lim_{N\rightarrow \infty}\frac{1}{N}\sum_{k=1}^N O[U^k, D^{-1}[U^k]] $.

The evaluation of form factors involves taking numerically the 
Fourier transform of two- and three-point functions
with respect to momentum transfer which, on a finite
box of spatial length $L$, takes
discrete values in units of $2\pi/L$. 
For large values of momentum transfer the results become noisy
and therefore we are limited
up to $Q^2\equiv -q^2$ $\sim 2$~GeV$^2$. 
To ensure that finite volume effects are
kept small we take box sizes such that  $Lm_\pi \stackrel{>}{\sim} 4.5$, where
$m_\pi$ is the pion mass\footnote{One exception is in the case of dynamical 
Wilson fermions at the smallest pion mass for which $Lm_\pi=3.6$
as  marked in the Table.}.
In addition, discretization errors due to the finite lattice spacing  $a$
must be checked. 
Wilson fermions have  ${\cal O}(a)$ discretization errors and
staggered fermions with  Asqtad action
and domain wall fermions (hybrid approach) have ${\cal O}(a^2)$ errors.
Therefore agreement between results in these two approaches
 provides an indication that cut-off effects are under control.
Finally, we use  larger  bare u- and d -quark masses than physical
 and extrapolation to the chiral limit must be considered.

Form factors are extracted from  three-point functions,
$G^{{\Delta} J N}(t_2,t_1;{\bf q})=<\Omega|\sum_{\bf x_1,x_2}
e^{i{\bf q}.{\bf x_1}}
\hat{T} \hat{J}_{\tilde{h}}({\bf x_2},t_2)\hat{J}({\bf x_1},t_1) \hat{J}_{h}^\dagger(0)|\Omega>$, shown in the diagram below:

    \begin{minipage}[h]{4cm}\hspace*{-0.3cm}
 \scalebox{0.45}{\begin{picture}(0,0)%
\includegraphics{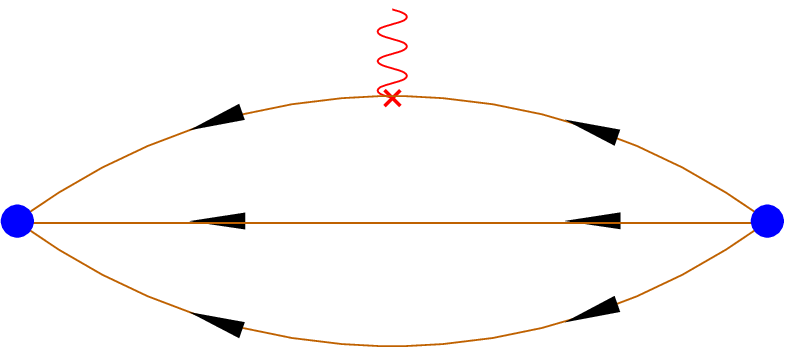}%
\end{picture}%
\setlength{\unitlength}{3947sp}%
\begingroup\makeatletter\ifx\SetFigFont\undefined%
\gdef\SetFigFont#1#2#3#4#5{%
  \reset@font\fontsize{#1}{#2pt}%
  \fontfamily{#3}\fontseries{#4}\fontshape{#5}%
  \selectfont}%
\fi\endgroup%
\begin{picture}(4133,1857)(218,-1636)
\put(250,-886){\makebox(0,0)[b]{\smash{\SetFigFont{12}{14.4}{\rmdefault}{\mddefault}{\updefault}{$\left(\vec{x}_2,t_2\right)$}%
}}}
\put(300,-1336){\makebox(0,0)[b]{\smash{\SetFigFont{12}{14.4}{\rmdefault}{\mddefault}{\updefault}{$\Delta\left(\vec{p}'\right)$}%
}}}
\put(2101,-736){\makebox(0,0)[b]{\smash{\SetFigFont{12}{14.4}{\rmdefault}{\mddefault}{\updefault}{$\left(\vec{x}_1,t_1\right)$}%
}}}
\put(4251,-1036){\makebox(0,0)[b]{\smash{\SetFigFont{12}{14.4}{\rmdefault}{\mddefault}{\updefault}{$\left(0,0\right)$}%
}}}
\put(4251,-1336){\makebox(0,0)[b]{\smash{\SetFigFont{12}{14.4}{\rmdefault}{\mddefault}{\updefault}{$N\left(\vec{p}\right)$}%
}}}
\put(2701,-286){\makebox(0,0)[b]{\smash{\SetFigFont{12}{14.4}{\rmdefault}{\mddefault}{\updefault}{$\vec{q}=\vec{p}'-\vec{p}$}%
}}}
\put(2101, 89){\makebox(0,0)[b]{\smash{\SetFigFont{12}{14.4}{\rmdefault}{\mddefault}{\updefault}{$J_\mu$}%
}}}
\end{picture}
}

  \end{minipage}
  \begin{minipage}[h]{6cm}\vspace*{0.3cm}
Interpolating fields for N and $\Delta$ are: 
    \beq \nonumber
    J^p(x)&=&\epsilon^{abc}[u^{Ta}(x)C\gamma_5d^b(x)]u^c(x),\\ \nonumber
    J^{\Delta^+}_\sigma(x)&=&\frac{1}{\sqrt{3}}\epsilon^{abc}\{2[u^{Ta}(x)C\gamma_\sigma d^b(x)]u^c(x)\\ \nonumber
      &+& [u^{Ta}(x)C\gamma_\sigma u^b(x)]d^c(x)\}
    \eeq
  \end{minipage}

\vspace*{0.3cm}

\noindent
In all cases we apply Gaussian smearing at the source and sink. 
In the case of  unquenched Wilson fermions  HYP-smearing is applied
to the gauge links used  in the Gaussian smearing of the source and sink. 
In the case of
domain wall fermions we  use HYP-smeared MILC configurations in all
computations.
We carry out  sequential inversions by fixing the quantum numbers at the
sink and source,
which means that the sink time $t_2$ is fixed, whereas 
the insertion time $t_1$ can vary and
 any operator can be inserted at $t_1$.
In this work we consider the vector current,
 $j_\mu^a=\bar{\psi}\gamma_\mu\frac{\tau^a}{2}\psi$,
the axial-vector current,
 $A_\mu^a=\bar{\psi}\gamma_\mu\gamma_5\frac{\tau^a}{2}\psi$
and the pseudoscalar current, $P^a=\bar{\psi}i\gamma_5\frac{\tau^a}{2}\psi$,
where $\tau^a$ are Pauli matrices acting in flavor space.
All $\vec{x}_1$ and $\vec{x}_2$ are summed over and we 
 vary $t_1$ in search for a plateau.
The exponential time  dependence 
and unknown overlaps of the interpolating fields with the 
physical states cancel by dividing the
three-point function with appropriate
combinations of two-point functions~\cite{Alexandrou:2004PRL}. 

The lattice parameters that we use are given in the Table.

\vspace*{-0.5cm}

\begin{center}
\scalebox{0.75}{
\begin{tabular}{|c|c|c|c|}
\hline \multicolumn{4}{c}{
{Wilson fermions}~~~}\\
\hline number of confs & $\kappa$ & $m_\pi$~(GeV) & $m_N$~(GeV) \\
\hline
\multicolumn{4}{|c|}{Quenched $32^3\times64$, $\beta=6.0,~~a^{-1}=2.14(6)$~GeV ($a=0.09$~fm) from nucleon mass at chiral limit} \\
\hline 200 &0.1554 &0.563(4) &1.267(11)\\
 200 &0.1558 &0.490(4)&1.190(13)\\
  200& 0.1562 &0.411(4) &1.109(13)\\
  &$\kappa_c$ =0.1571& 0.& 0.938(9)\\
  \hline
  \multicolumn{4}{|c|}{Unquenched~\cite{TchiL}  $24^3\times40$,$\beta=5.6,~~a^{-1} = 2.56(10)$~GeV ($a=0.08$~fm)}\\
  \hline
  185 &0.1575 &0.691(8) &1.485(18)\\
  157 &0.1580 &0.509(8) &1.280(26)\\
  \hline
  \multicolumn{4}{|c|}{Unquenched~\cite{Urbach} $24^3\times32$,$\beta=5.6,~~a^{-1} = 2.56(10)$~GeV }\\
  \hline
  200 & {0.15825} &{0.384(8)}{$\leftarrow Lm_\pi=3.6$} &1.083(18) \\
  &$\kappa_c$ = 0.1585 &0. &0.938(33)\\
  \hline
\end{tabular}
}
\end{center}

\begin{center}
\scalebox{0.75}{
\begin{tabular}{ccccccc}
\hline \multicolumn{7}{c}{
{Hybrid scheme}~~~$a^{-1}=1.58$~GeV ($a=0.125$~fm) from MILC collaboration}\\
\hline number of confs&Volume & $(am_{u,d})^{\textrm{sea}}$ &
$(am_{s})^{\textrm{sea}}$ & $(am_q)^{DW}$ & $m^{DW}_{\pi}$ (GeV) &
$m_N$ (GeV) \\
\hline
150&$20^3\times64$ & 0.03 &0.05 &0.0478&0.589(2)& 1.392(9)\\
198&$20^3\times64$ & 0.02 &0.05 &0.0313& 0.501(4)& 1.255(19)\\
100&$20^3\times64$ & 0.01 & 0.05 &0.0138 & 0.362(5) & 1.138(25)\\
150(300 for CMR) &$28^3\times64$& 0.01& 0.05 &0.0138& 0.354(2) & 1.210(24) \\
\hline
\end{tabular}
}
\end{center}

\section{$N$ to $\Delta$ electromagnetic form factors}\label{sec:EM NtoDelta}
The N to $\Delta$ matrix element of the electromagnetic current
can be decomposed into a
 dominant magnetic dipole, $G_{M1}$, and two suppressed electric and  Coulomb 
quadrupole form
factors, $G_{E2}$ and $G_{C2}$. A non-zero
 $G_{E2}$ and  $G_{C2}$  
signal a deformation in the nucleon and/or $\Delta$.
Precise experimental data on the quadrupole  to dipole ratios,
 $ R_{EM}({\rm EMR)} = -\frac{G_{E2}(q^2)}{G_{M1}(q^2)}\>,$ and
 $ R_{SM}({\rm CMR}) = -\frac{\vert\vec{q}\vert}{2m_\Delta}
  \frac{G_{C2}(q^2)}{G_{M1}(q^2)},$
suggest deformation of the nucleon/$\Delta$~\cite{papanicolas:2003}.


\begin{figure}
 \vspace*{-0.5cm}
\begin{minipage}{5.5cm}
\psfig{file=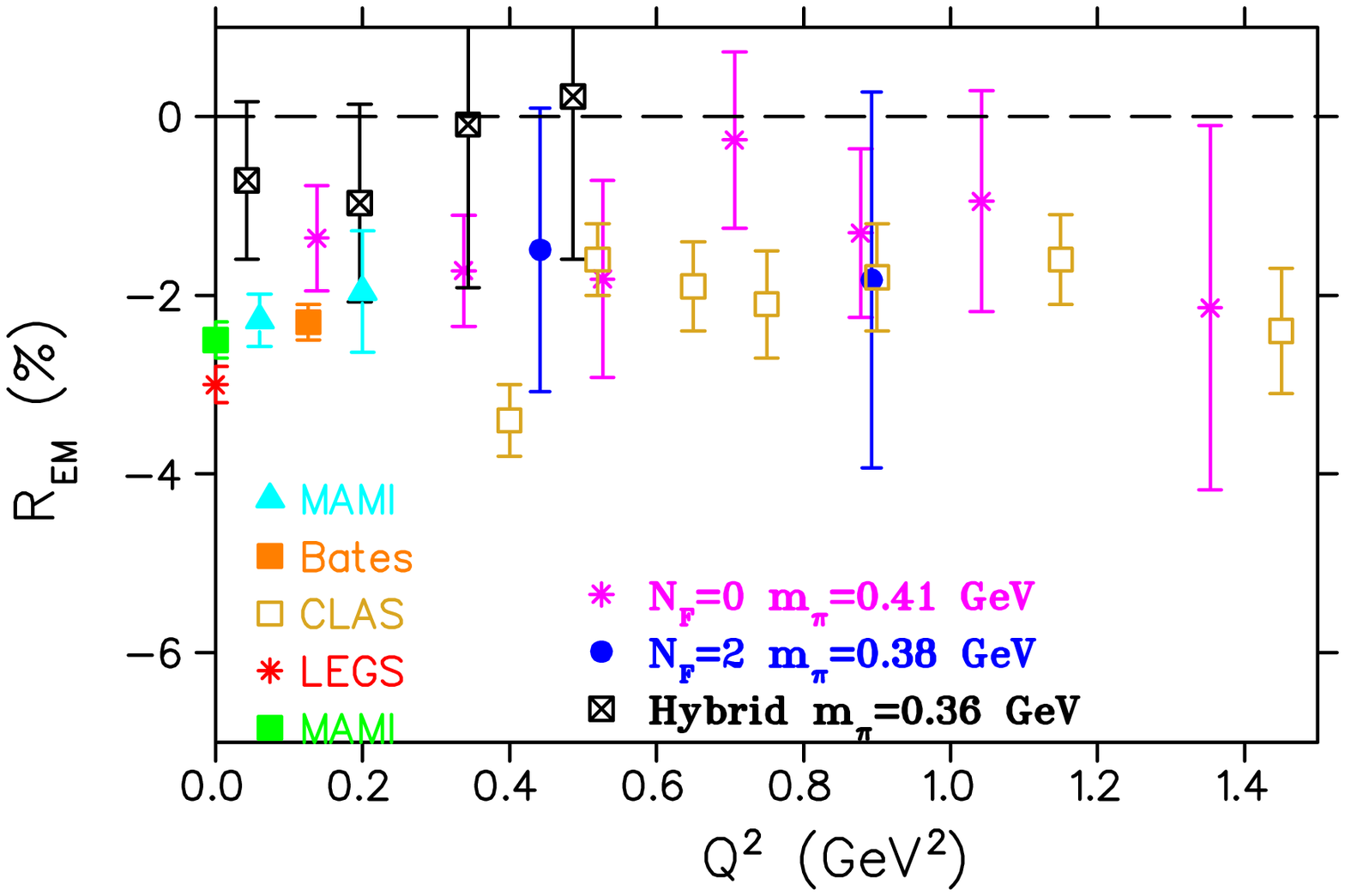,width=5.5cm}
\end{minipage}
\begin{minipage}{5.5cm}
\psfig{file=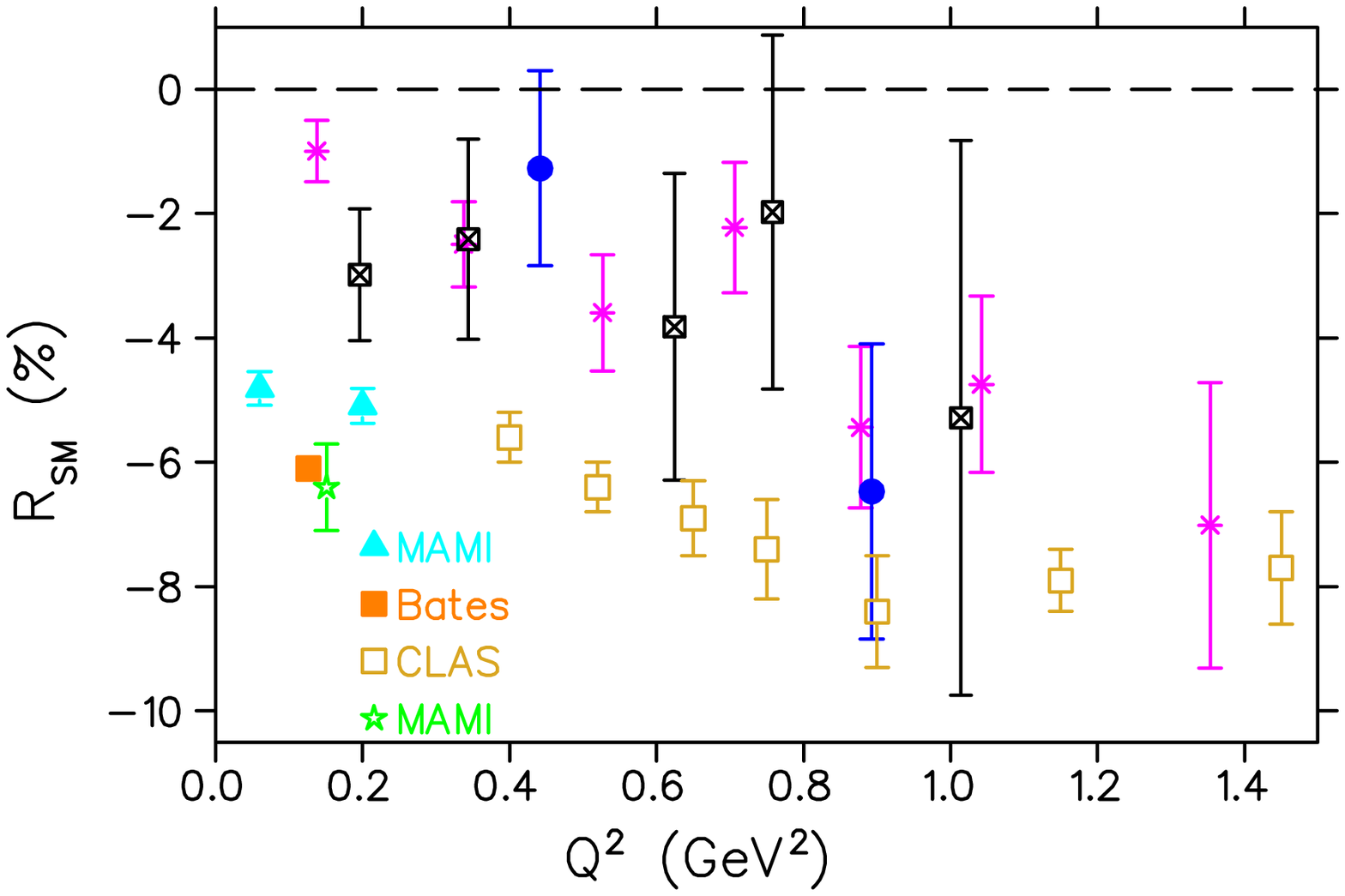,width=5.5cm}
\end{minipage}
\caption{The EMR (left) and CMR (right) for the lightest 
pion mass in our three type of simulations. }
\label{fig:EMR-CMR}
\vspace*{-0.3cm}
\end{figure}

In Fig.~\ref{fig:EMR-CMR}  we show the EMR and CMR ratios
for the smallest pion mass in the quenched case, for two dynamical
flavors of Wilson fermions and in the hybrid approach.
For the first time in full QCD, we achieve good enough accuracy to exclude
a zero value for these ratios. Furthermore, at low $Q^2$, unquenched results
become more negative bringing lattice results closer to
experiment and showing the importance of the pion cloud effects
at small $Q^2$.

\begin{figure}
\begin{minipage}{5.5cm}
\psfig{file=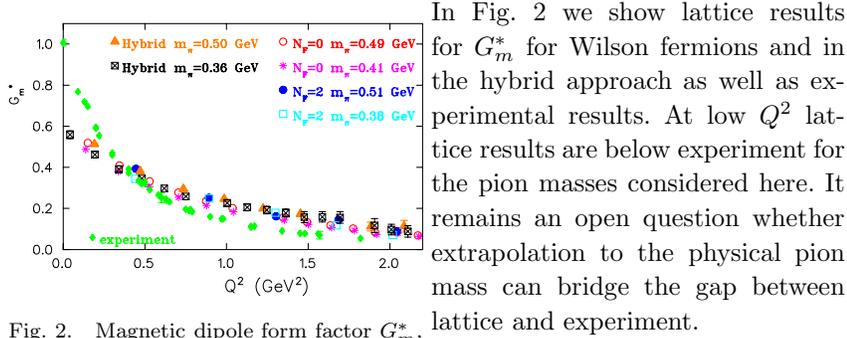,width=5.5cm}
\caption{Magnetic dipole form factor $G_m^*$, in the Ash parameterization:
$G^*_m=\frac{1}{\sqrt{1+\frac{Q^2}{(m_N+m_\Delta)^2}}} G_{M1}$.}
\label{fig:GM1}
\end{minipage}
\begin{minipage}{5.5cm}\vspace*{-1.5cm}
In Fig.~\ref{fig:GM1} we show lattice results for $G_m^*$
for Wilson fermions and in the hybrid approach  as well as experimental 
results. 
At low
$Q^2$ lattice results are below experiment for the pion masses
considered here. It remains an open question whether extrapolation to the
physical pion mass  can bridge the gap between 
lattice and experiment.
\end{minipage}
\vspace*{-1.3cm}
\end{figure}

\section{Nucleon and $N$ to $\Delta$ axial-vector form factors}
In the case of Wilson
fermions, besides N to $\Delta$ 
we also calculate the nucleon axial-vector  form factors.  
 The LHP 
collaboration~\cite{LHPC07}
has evaluated these form factors
 in the hybrid approach with the same parameters
as those used in our N to $\Delta$ study and therefore, in
this case,  we use their results
to compare. 
The  nucleon axial- vector form factors $G_A$ and $G_p$ are given by
\small
\be
\langle N(p')|A_\mu^3|N(p)\rangle=i \sqrt{
            \frac{m_N^2}{E_N({\bf p}')E_N({\bf p})}} 
            \bar{u}(p') \Bigg[
            G_A(q^2)\gamma_\mu\gamma_5 
 +\frac{q_\mu \gamma_5}{2m_N}G_p(q^2)\Bigg]\frac{\tau^3}{2}u(p)
\ee 
\normalsize
Since the final state is no longer the $\Delta$ a new set of sequential
inversions is needed. 
The  decomposition of the N to $\Delta$  matrix element of the axial-vector
current can be written 
in terms of four transition form factors~\cite{Adler}:
\small
\beq
<\Delta(p^{\prime},s^\prime)|A^3_{\mu}|N(p,s)> &=& 
i\sqrt{\frac{2}{3}}  \sqrt{
            \frac{m_N m_\Delta}{E_\Delta({\bf p}')E_N({\bf p})}}\>\>
\bar{u}_{\Delta^+}^\lambda(p^\prime,s^\prime)\biggl[C^A_5(q^2) g_{\lambda\mu}  \nonumber \\
&\>&\hspace*{-4.1cm}+\frac{C^A_6(q^2)}{m^2_N} q_\lambda q_\mu +\left (\frac{C^A_3(q^2)}{m_N}\gamma^\nu + \frac{C^A_4(q^2)}{m^2_N}p{^{\prime \nu}}\right)  
\left(g_{\lambda\mu}g_{\rho\nu}-g_{\lambda\rho}g_{\mu\nu}\right)q^\rho \biggr]
u_P(p,s)
\label{ND axial}
\eeq
\normalsize
  Under the assumptions that { $C^A_3\sim 0$} and
    { $C^A_4 \ll C^A_5$} the parity violating asymmetry is
    proportional to the ratio  $C^A_5/C^V_3$~\cite{Nimai:1998}, 
 where  $C^V_3$ can be obtained from the electromagnetic N to
    $\Delta$ transition.

   \begin{figure}
\begin{minipage}{5.7cm}
 \epsfxsize=5.5truecm \epsfysize=4truecm
   \mbox{\epsfbox{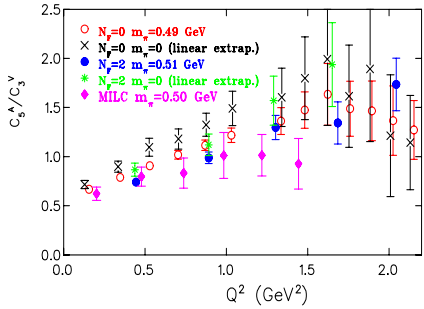}}
\caption{Parity violating asymmetry.}
\label{fig:parity}
\end{minipage}
\begin{minipage}{5.5cm}\vspace*{-0.8cm}
As shown in Fig.~\ref{fig:parity},  
the asymmetry is non-zero when $Q^2=0$~\cite{Alexandrou:2004PRL},
increases with $Q^2$ up to about $Q^2\sim 1.5$~GeV$^2$
and shows small unquenching effects  for this range of quark masses.
Given this  weak quark mass dependence, the results can be taken as 
a prediction for the ratio to be measured by the G0 collaboration~~\cite{G0:2002}.
\end{minipage}
\vspace*{-0.5cm}
   \end{figure}

\vspace*{-0.5cm}

\subsubsection{Goldberger-Treiman relations}
\begin{figure}
 
\begin{minipage}{5.5cm}
\psfig{file=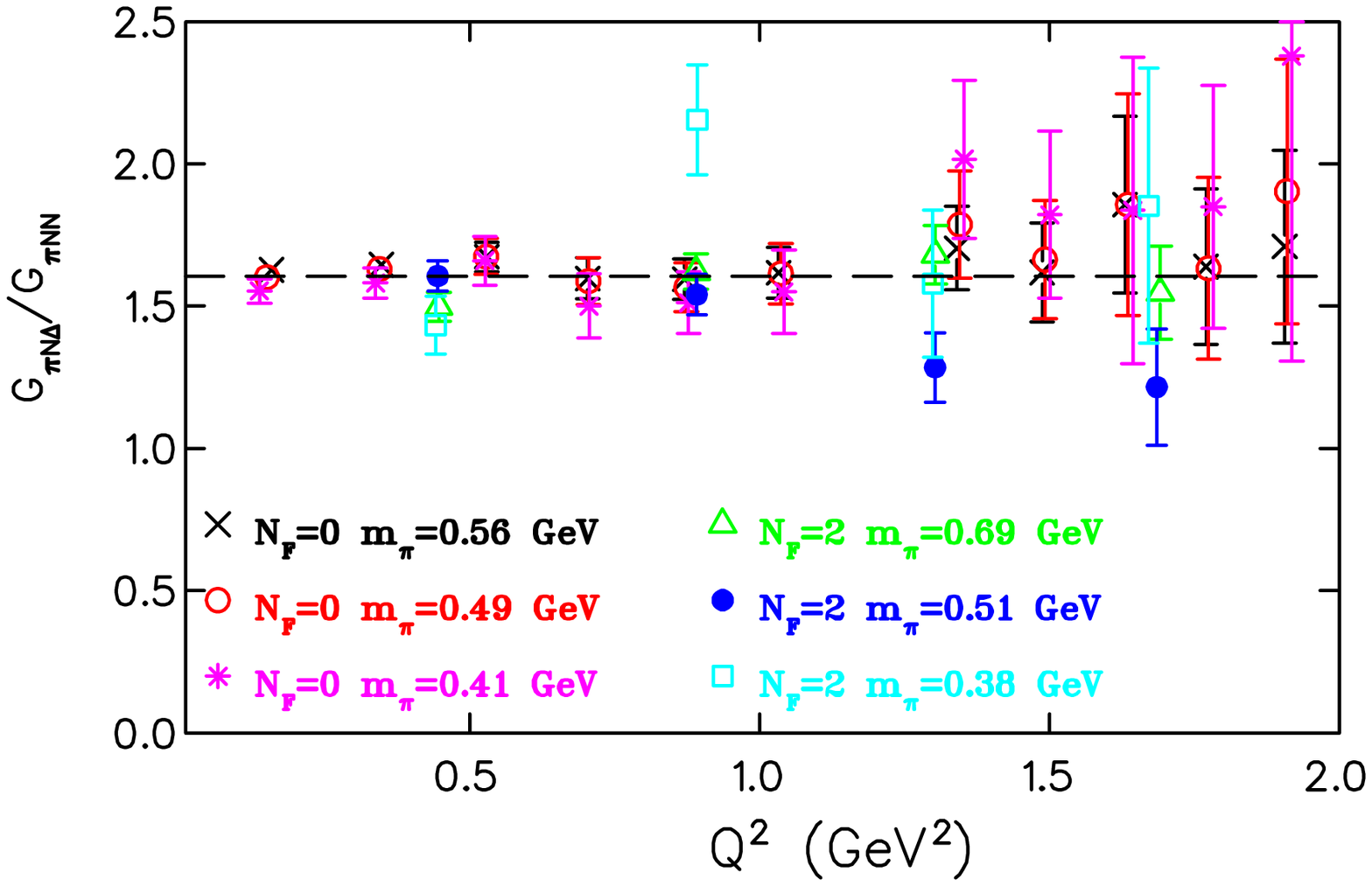,width=5.5cm}
\end{minipage}
\begin{minipage}{5.5cm}
\psfig{file=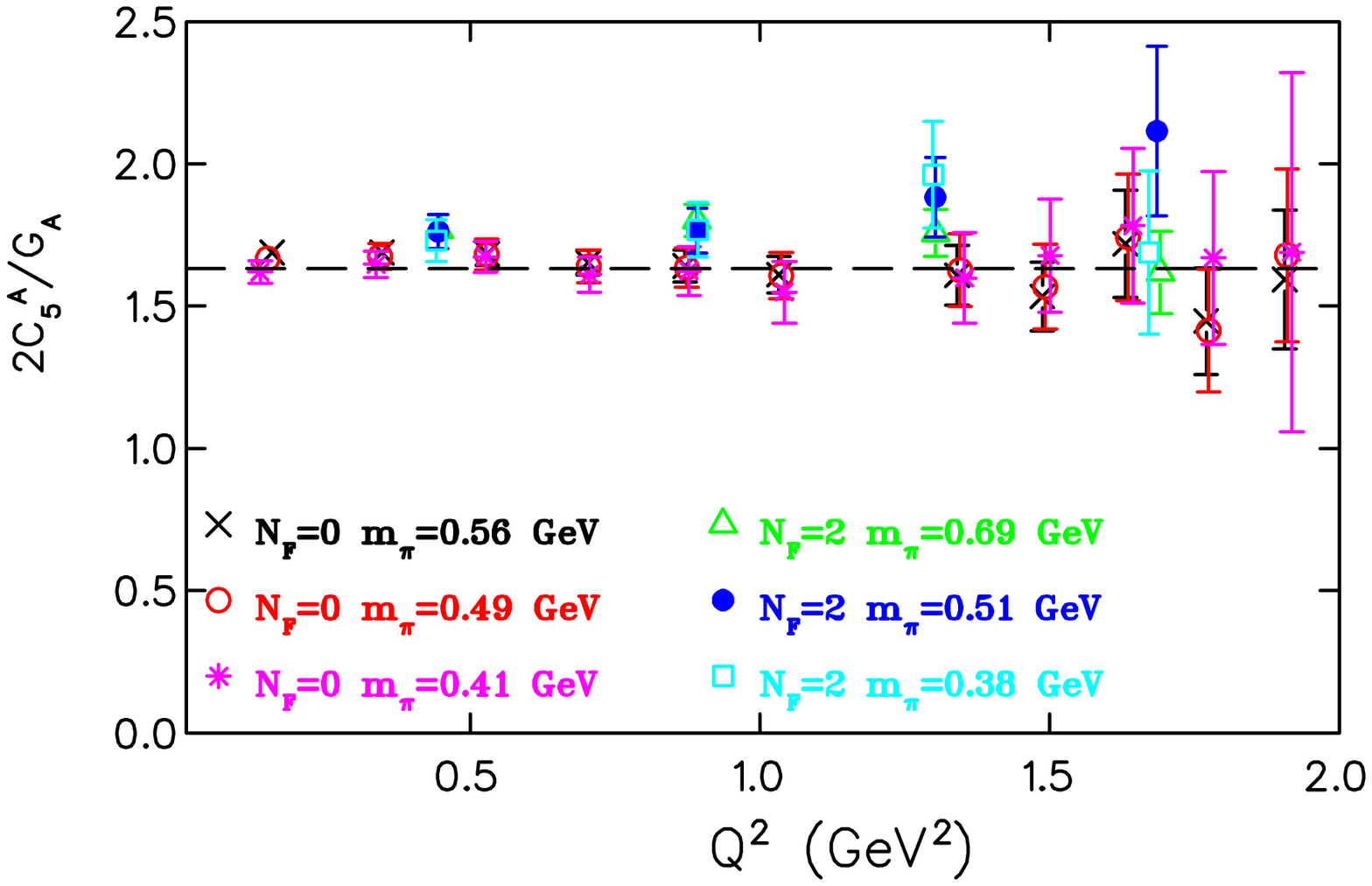,width=5.5cm}
\end{minipage}
\caption{The ratios ${G_{\pi N \Delta}}/G_{\pi NN}$ (left) and
$2C_5^A/G_A$ (right).}
\label{fig:GpiNDoverGpiNN}
\vspace*{-0.5cm}
\end{figure}

 Partial conservation of axial current (PCAC),
$\partial^\mu A_\mu^a=f_\pi m_\pi^2 \pi^a$,
and the axial Ward Identity, 
$ \partial^\mu A_\mu^a=2m_qP^a$,
 relate the pion field $\pi^a$ with the 
pseudoscalar density: $\pi^a=\left({2m_qP^a}/{f_\pi m_\pi^2}\right)$,
where the pion decay constant $f_\pi$ is determined from the two-point function
$ <0| A^{a}_{\mu} | \pi^{b} (p) > = i p_\mu \delta^{ab} f_{\pi}$.
The renormalized quark mass, $m_q$, is given by
$m_q = \frac{m_\pi<0|\tilde{A}_0^a|\pi^a(0)>}{2<0|\tilde{P}^a|\pi^a(0)>}$,
where  $\tilde{A}_0^a$ and $\tilde{P}^a$ are the renormalized currents. 
To obtain  the $\pi NN$ and  $\pi N\Delta$  form factors we use the decomposition
\small
\beq
 2m_q\langle N(p^\prime)|P^3|N(p)\rangle&=& 
i\sqrt{\frac{m_N^2}{E_N({\bf p}')E_N({\bf p})}}
\frac{f_\pi m_\pi^2\> G_{\pi NN}(q^2)}
{m_\pi^2-q^2}\>\bar{u}(p^\prime)\gamma_5 \frac{\tau^3}{2}u(p) \nonumber \\
 2m_q\langle\Delta(p^\prime)|P^3|N(p)\rangle &=& 
i\sqrt{\frac{2}{3}}
\sqrt{\frac{m_\Delta m_N}{E_\Delta({\bf p}^\prime) E_N({\bf p})}}
\frac{f_\pi m_\pi^2 \>G_{\pi N\Delta}(q^2)}
{m_\pi^2-q^2}
\bar{u}_{\Delta^+}^\nu(p^\prime) \frac{q_\nu}{2m_N}u_P(p) \nonumber
\label{Gpi}
\eeq
\normalsize
PCAC relates the axial form factors $G_A$ and $G_p$ 
with $G_{\pi NN}$ 
and equivalently $C_5^A$ and $C_6^A$ with $G_{\pi N\Delta}$. These
are the well known generalized Goldberger-Treiman relations (GTRs). 
As mentioned above there are advantages in  considering ratios. In
Fig.~\ref{fig:GpiNDoverGpiNN}  we show two such ratios,
namely ${G_{\pi N \Delta}}/{G_{\pi NN}}$ and $2C_5^A/{G_A}$.
Both are
 independent of $Q^2$ and the quark mass. Fitting to a constant
we find  ${2C_5^A}/{G_A}\sim 1.6\sim{G_{\pi N \Delta}}/{G_{\pi NN}}$,
which
implies the Goldberger-Treiman relations: 
$G_{\pi NN}(q^2)\>f_\pi =m_N G_A(q^2)$ and 
$G_{\pi N \Delta}(q^2)\>f_\pi = 2m_N C_5^A(q^2).$
Assuming pion-pole dominance we can write
$G_p(Q^2)=\frac{4m_N^2/m_\pi^2}{1+Q^2/m_\pi^2}\> G_A(Q^2)$ and
$C_6^A(Q^2)=\frac{m_N^2/m_\pi^2}{1+Q^2/m_\pi^2}\> C_5^A(Q^2).$
Therefore we have the equality
$8C_6^A(Q^2)/G_p(Q^2)={G_{\pi N \Delta}}/G_{\pi NN}$.
We find that $8C_6^A(Q^2)/G_p(Q^2) \sim 1.7$~\cite{Alexandrou:2007prd}
a few percentage larger than
${G_{\pi N \Delta}}/{G_{\pi NN}}$. 

\begin{figure} 
\begin{minipage}{5.5cm}
\psfig{file=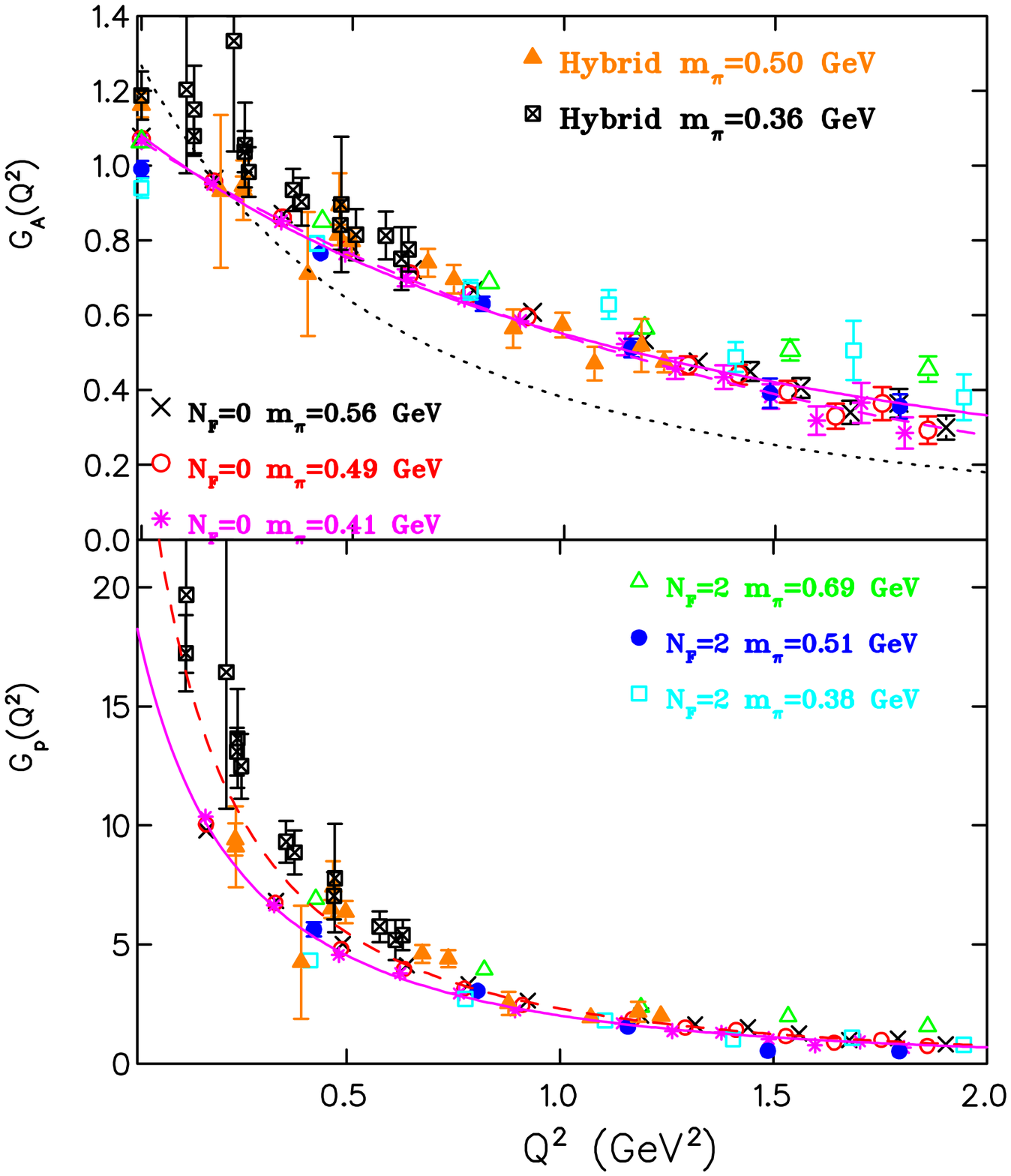,width=5.5cm}
\end{minipage}
\begin{minipage}{5.5cm}
\psfig{file=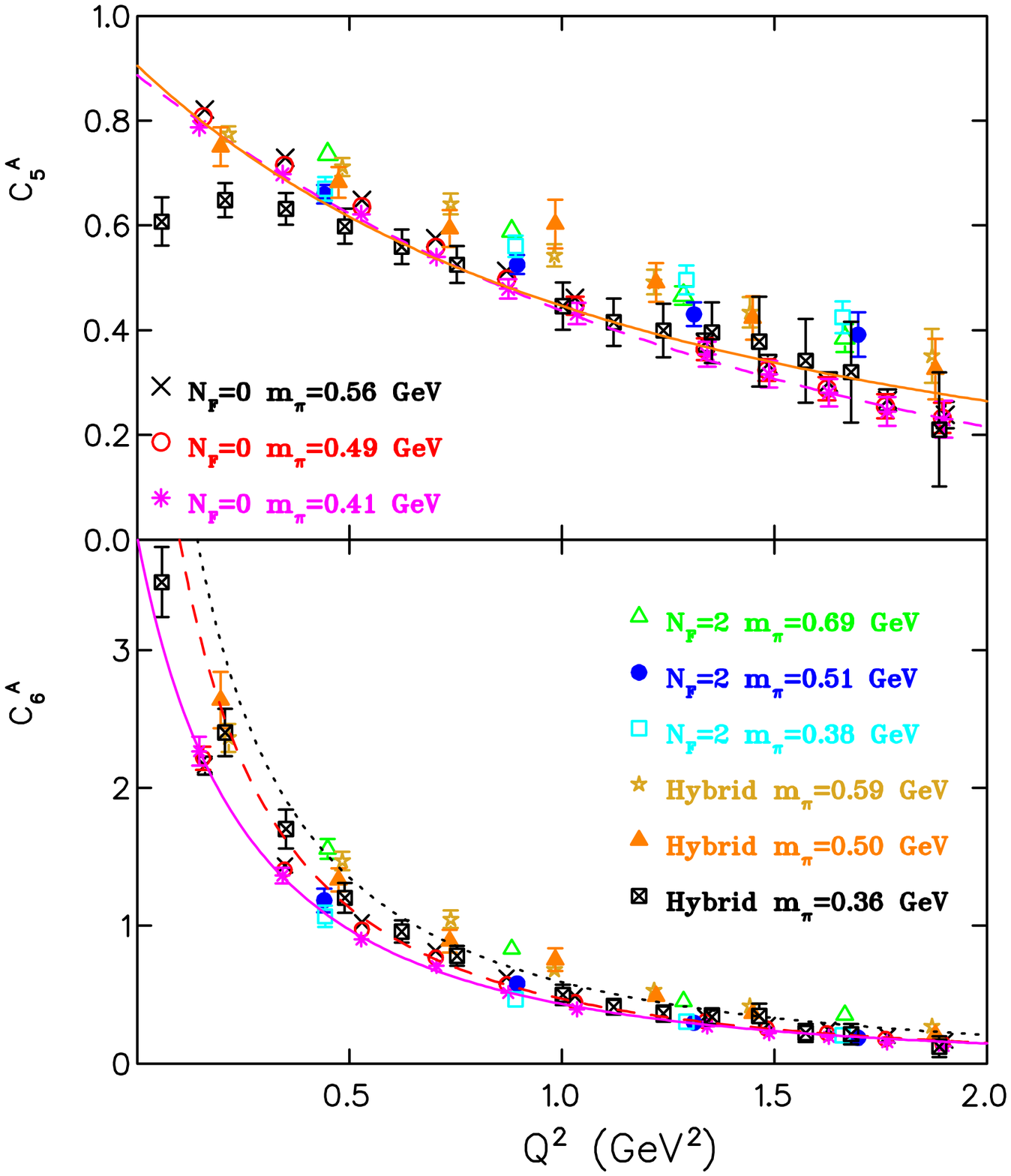,width=5.5cm}
\end{minipage}
\caption{The nucleon axial form factors $G_A$ (top, left) and
$G_p$ (bottom, left) and the
N to $\Delta$ axial form factors  $C^A_5$ (top, right) and $C^A_6$ (bottom, right).}
\label{fig:GAHA}
\vspace*{-0.5cm}
\end{figure}

In Fig.~\ref{fig:GAHA}  we present
the nucleon and N to $\Delta$ axial form factors separately 
together with fits of $G_A$ and $C_5^A$ to a dipole form,
$g_0/(\frac{Q^2}{m_A^2}+1)^2$.
Dynamical QCD results in the hybrid approach for
the smallest pion mass, where we can access low
$Q^2$ values, show large unquenching effects.
Having fitted  $G_A$ and $C_5^A$, we  
 can check if pion-pole dominance describes 
the $Q^2$-dependence of   $G_p$ and $C_6^A$. 
 The dashed lines correspond to the quenched data and show the behavior
for $G_p$ and $C_6^A$ extracted from fits
to $G_A$ and $C_5^A$ assuming pion-pole dominance, 
 whereas the dotted line shown for  $C_5^A$ is
for the hybrid approach, in both cases for the lightest pion mass.
As can be seen, they deviate from the lattice results at low $Q^2$.
Instead they are best described by the 
solid curves, which are obtained by fitting the pole mass.

\begin{figure}[h]
\begin{minipage}{5.5cm}
\epsfxsize=5.5truecm \epsfysize=7truecm
\mbox{\epsfbox{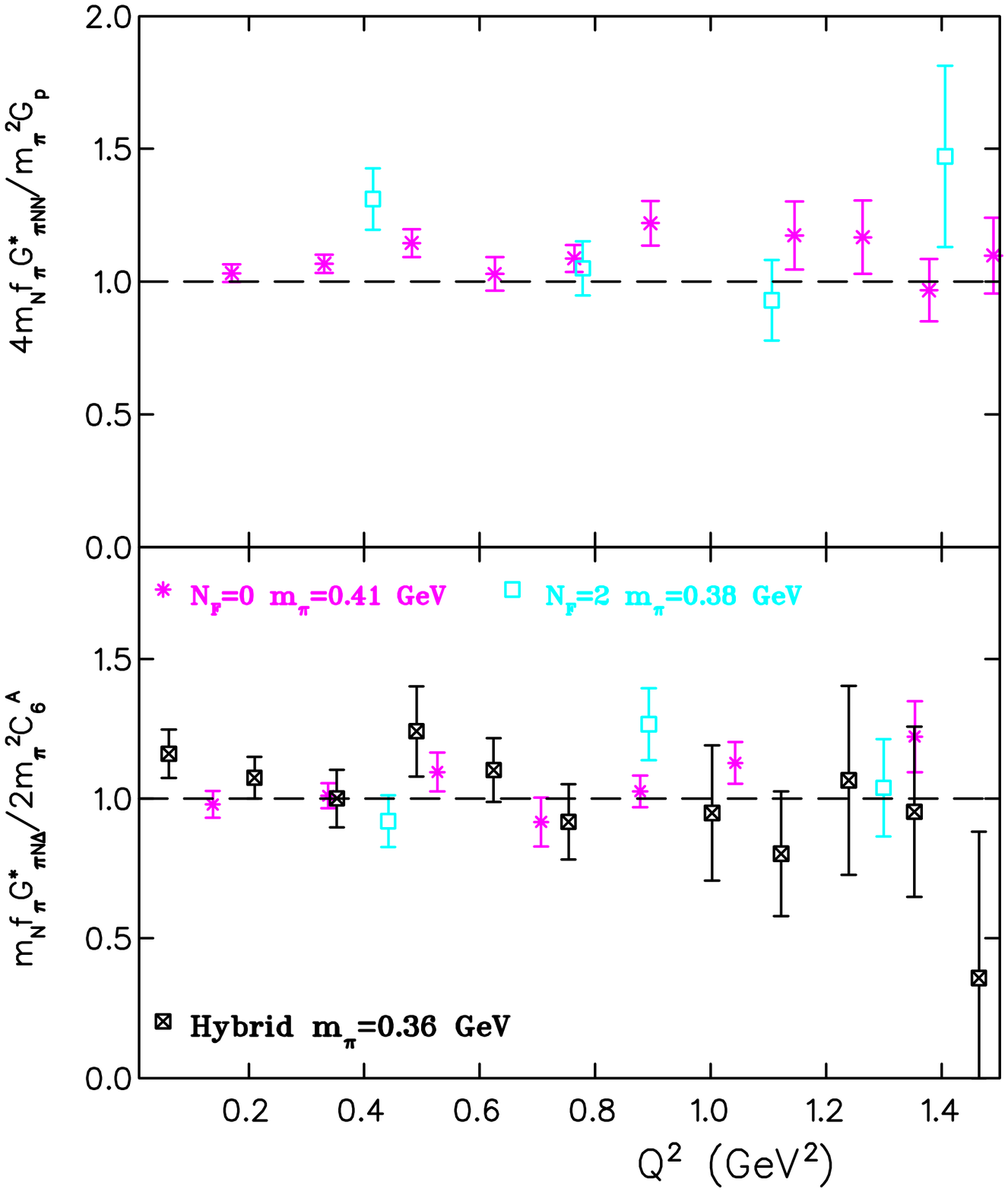}}
\end{minipage}
\begin{minipage}{5.5cm}
\epsfxsize=5.5truecm \epsfysize=7truecm
\mbox{\epsfbox{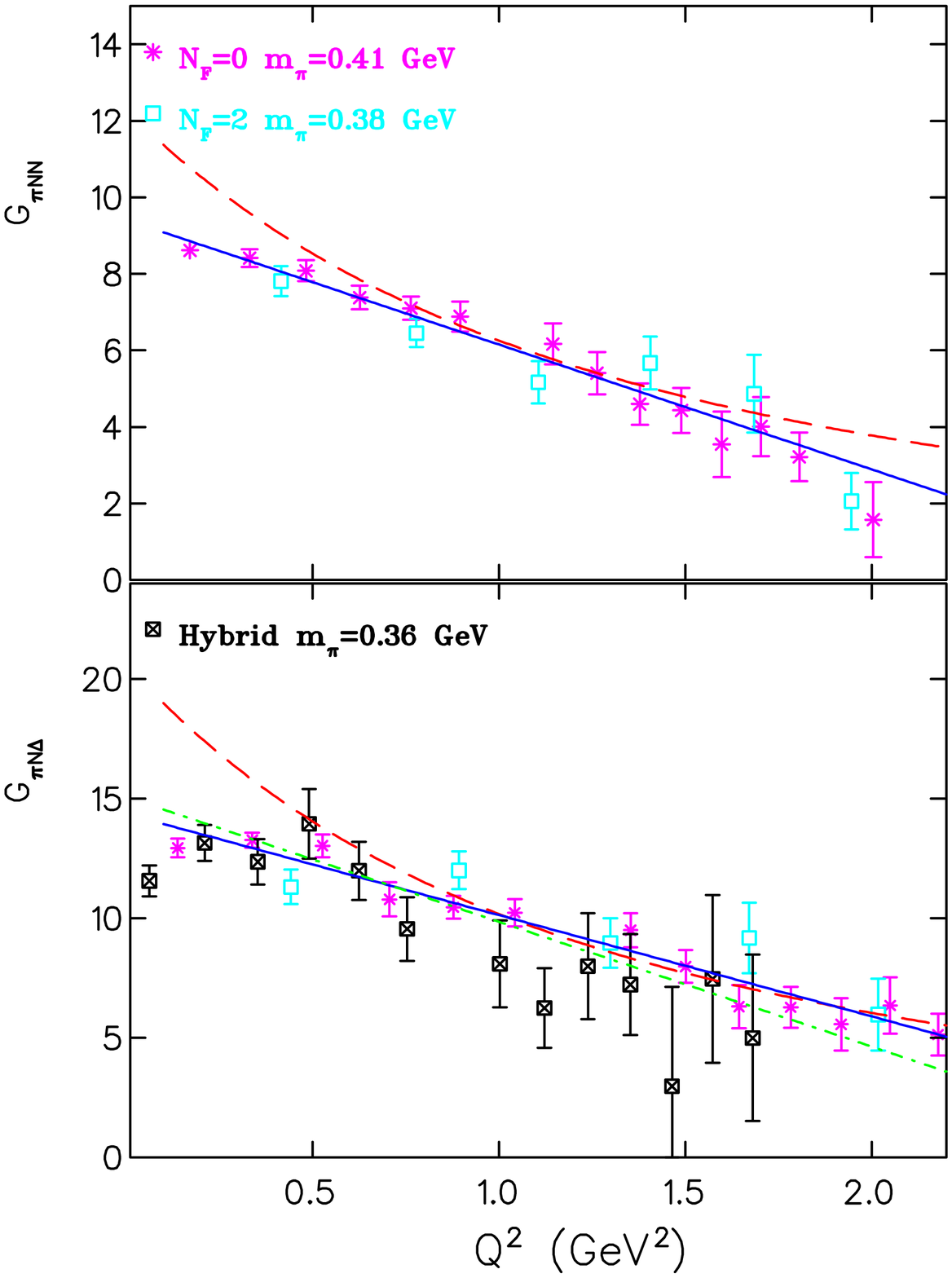}}
\end{minipage}
\caption{Left: The ratios  $R_{NN}$ (top) and $R_{N\Delta}$ (bottom)
and right: $G_{\pi NN}$ (top) and $G_{\pi N\Delta}$ (bottom) 
 for the smallest pion mass
in each type of simulation.}
\label{fig:Gpis}
\vspace*{-0.3cm}
\end{figure}
In Fig.~\ref{fig:Gpis} we show the relations
\small 
\be R_{NN}\equiv \frac{4m_Nf_\pi G^*_{\pi NN}(Q^2)}{m_\pi^2G_p(Q^2)}, ~~~~~~
R_{N\Delta}\equiv \frac{m_Nf_\pi G^*_{\pi N\Delta}(Q^2)}{2m_\pi^2C_6^A(Q^2)}, 
\ee
\normalsize
where we have defined 
$ G^*_{\pi NN}(Q^2)\equiv G_{\pi NN}(Q^2)/(1+Q^2/m_\pi^2)$
with a corresponding expression for $G^*_{\pi N\Delta}$.
As can be seen these ratios are consistent with unity for all $Q^2$-values. 
Finally, in Fig.~\ref{fig:Gpis} we show $G_{\pi NN}$ and $G_{\pi N\Delta}$
for the smallest pion mass.
The dash lines
 are obtained from fits of  $G_A$ and $C_5^A$ 
via the GTRs, $G_{\pi NN}(Q^2)=  m_N G_A(Q^2)/f_\pi$ and 
$G_{\pi N \Delta}(Q^2)= 2m_N C_5^A(Q^2)/f_\pi$.
As can be seen,
 there are large deviations 
at small  $Q^2$.
Lattice results at this pion mass give a smaller value in the limit 
$Q^2\rightarrow 0$ than what is extracted from experimental data namely,
$G_{\pi NN}(0)=13.21(11)$~\cite{exper:gpiNN}. The solid lines
are fits to the form 
$\biggl(1 -\Delta\frac{Q^2}{m_\pi^2}\biggr)$
with $a, \Delta$ fit parameters. Using these fits 
 $G_{\pi NN}(0)$ and $G_{\pi NN}(0)$ are extracted
(for details see Ref.~\cite{Alexandrou:2007prd}).

\vspace*{-0.2cm}

\section{Conclusions}
Lattice results on the electromagnetic, axial-vector and pseudoscalar 
form factors for the nucleon and the  N to $\Delta$ transition  
are presented in the quenched approximation, for two-flavors
 of dynamical Wilson fermions and using dynamical staggered
sea quarks and domain wall valence quarks (hybrid approach). Results 
 on the  quadrupole to dipole ratios EMR and CMR, obtained in 
the hybrid approach reaching down to a pion mass of 350 MeV and low
$Q^2$-values, are non-zero and of similar magnitude as in experiment.
We also find that ratios of form factors, such as 
 $G_{\pi N\Delta}/G_{\pi NN}\sim 1.6$
and $2C^A_5/G_A\sim 1.6$, calculated using Wilson
fermions, are in agreement with phenomenology.
Our  results for  the ratio {$C^A_5/C^V_3$} as a function
of $Q^2$, can be regarded as a lattice prediction 
for the parity violating asymmetry to leading order.
The deviations from experiment seen for the magnetic 
dipole N-$\Delta$ transition form factor $G_m^*$ and the values
of $G_{\pi NN}$ and
$G_{\pi N\Delta}$ in the limit $Q^2\rightarrow 0$ need further study.
In particular, finite lattice spacing effects,
as well as, chiral extrapolation to the physical pion mass must be 
investigated.

\vspace*{0.3cm}

\noindent
{\bf Acknowledgments:} I would like to thank my collaborators 
G. Koutsou, Th. Leontiou, 
H. Neff, J. W. Negele,  W. Schroers and A. Tsapalis for their
valuable contributions that made this work possible. This work is
supported in part by the  EU Integrated Infrastructure Initiative
Hadron Physics (I3HP) under contract RII3-CT-2004-506078.

\vspace*{-0.3cm}
\bibliographystyle{ws-procs9x6}
\bibliography{TransitionFF}


\end{document}